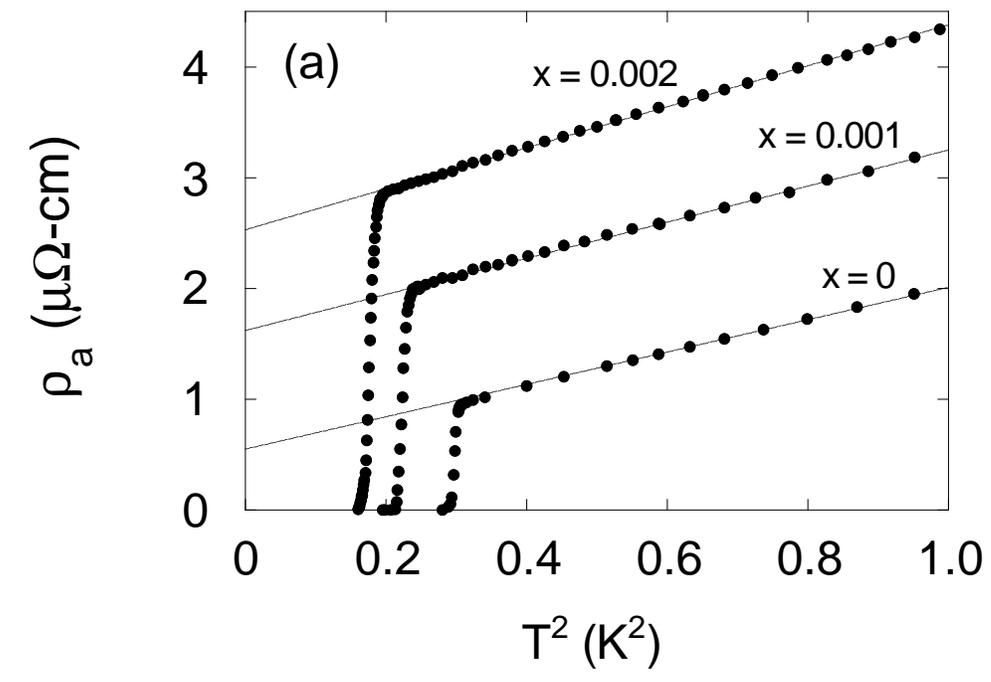

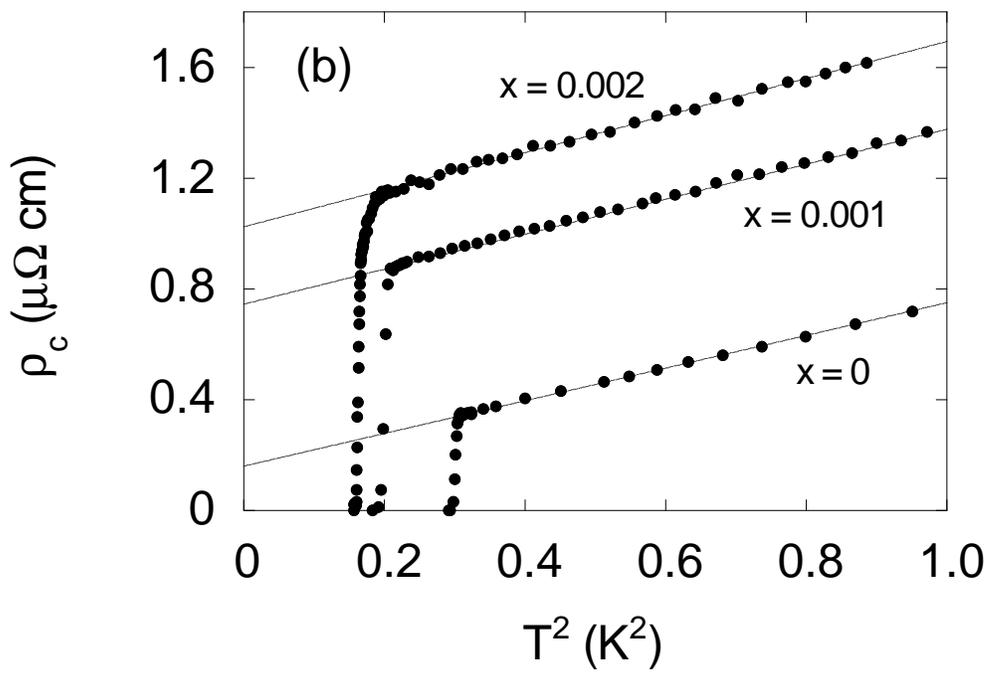

**Fig. 1**

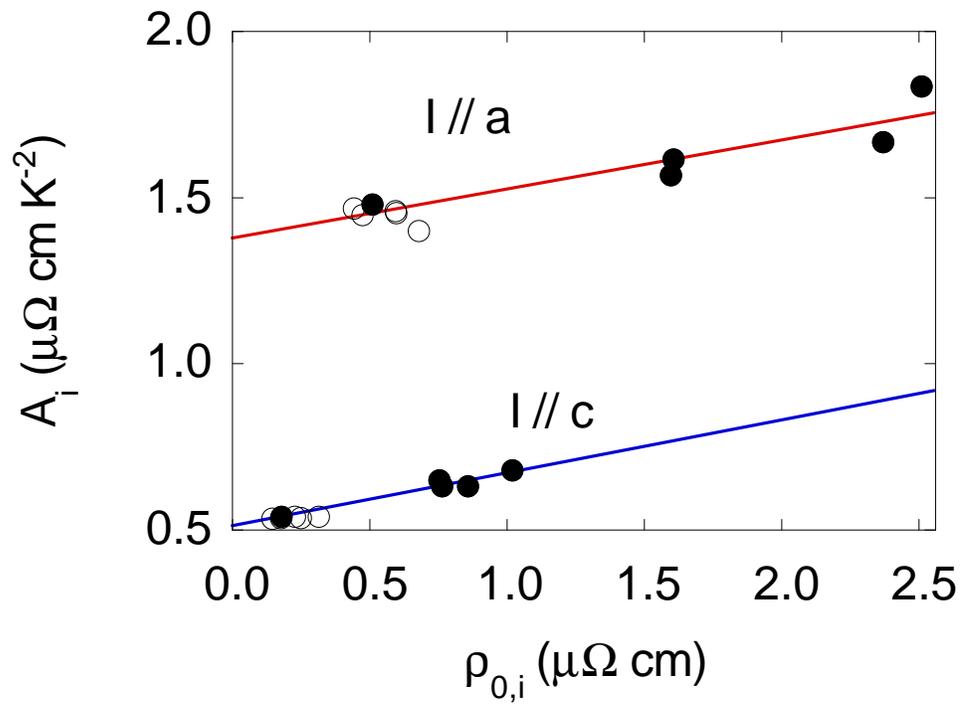

**Fig. 2**

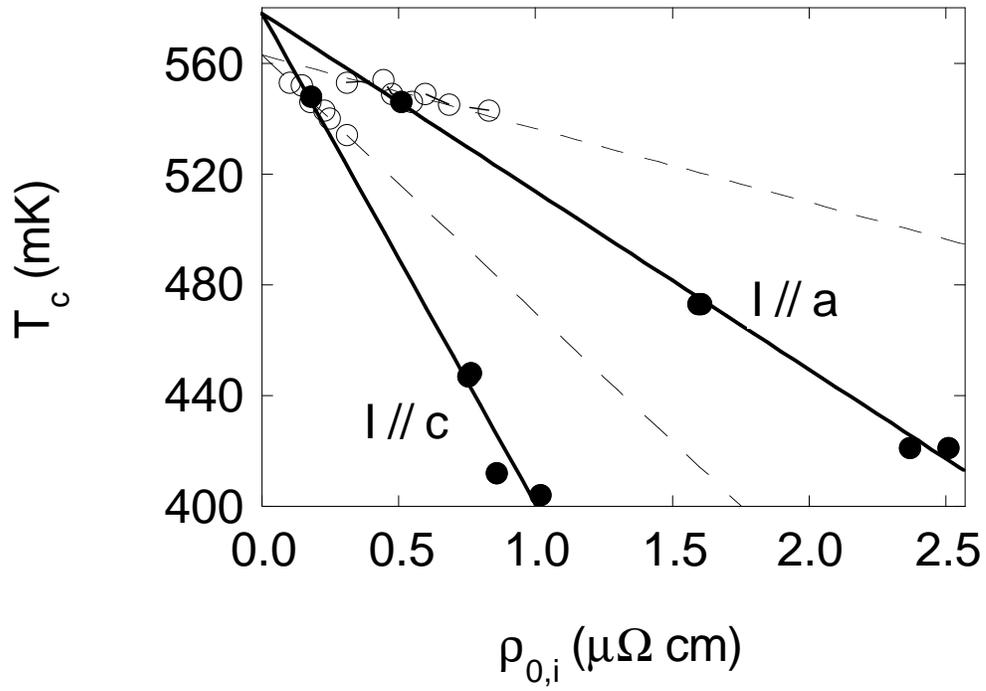

**Fig. 3**

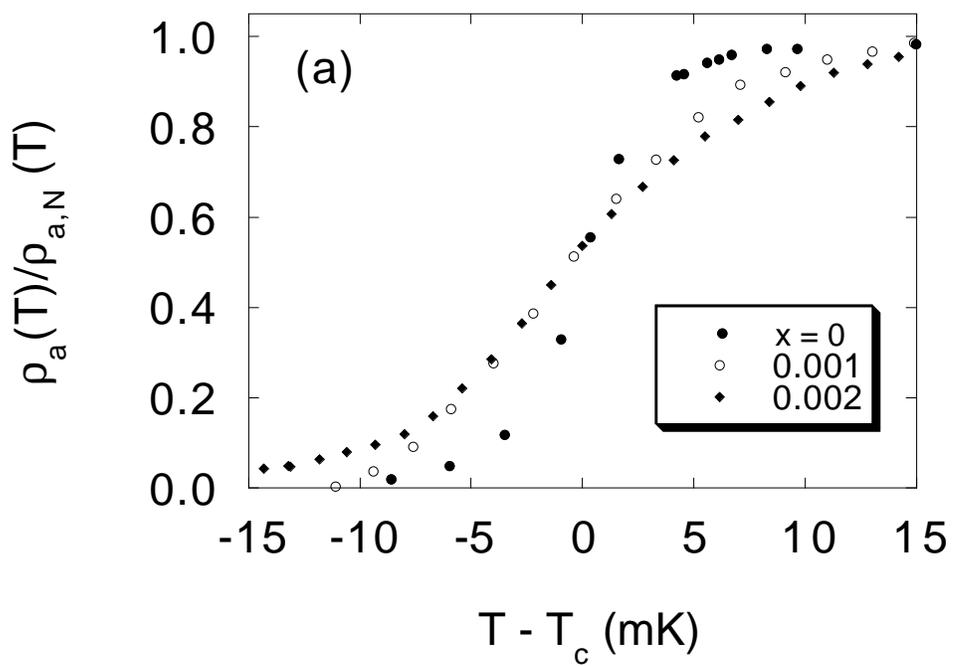

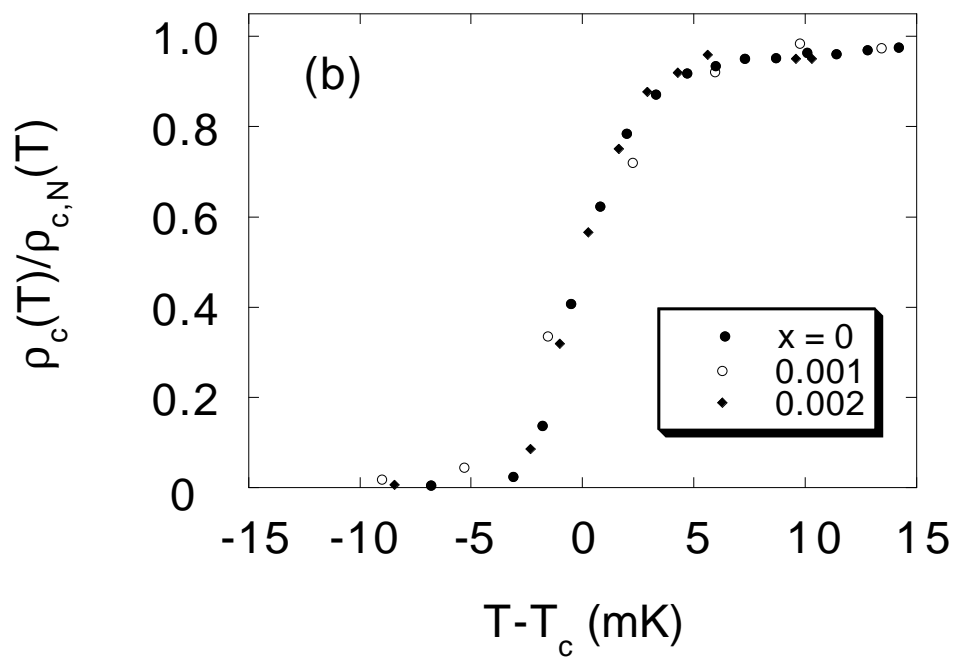

**Fig. 4**

# Suppression of Superconductivity in Single Crystals of UPt$_3$ by Pd Substitution

M. J. Graf

*Department of Physics, Boston College, Chestnut Hill, MA 02467 USA*

R. J. Keizer, A. de Visser, A. A. Menovsky, and J. J. M. Franse

*Van der Waals – Zeeman Institute, University of Amsterdam, 1018XE Amsterdam, The Netherlands*

The suppression of superconductivity by substitution effects has been measured in high quality single crystals of U(Pt$_{1-x}$ Pd$_x$)$_3$ with $0 \leq x \leq 0.002$. While the superconducting transition temperature T$_c$ varies linearly with residual resistivity $\rho_0$, consistent with pair-breaking by impurity potential scattering, the rate of suppression of T$_c$ with $\rho_0$ is much larger for Pd substitution than for other impurity substitutions or by increased defect density. This effect is correlated with an increase in the inelastic scattering coefficient, and may be related to Pd-induced changes in the magnetic fluctuation spectrum.

Pd-substitution is a powerful method to study unconventional superconductivity and magnetism in UPt$_3$. Early studies of polycrystalline U(Pt$_{1-x}$Pd$_x$)$_3$ for x ≤ 0.002 showed that Pd-substitution is unique in that it increases the splitting $\Delta T_c$ between transition temperatures of the A and B superconducting phases by 40 mK for x=0.002.[1] Addition of Pd also increases the zero-temperature moment associated with the anomalous "small-moment" antiferromagnetic phase (SMAF) above the x = 0 value of 0.02 $\mu_B$, and the correlation between the increased moment and $\Delta T_c$ has been confirmed.[2] For x > 0.01, "large-moment" antiferromagnetism (LMAF) is observed through neutron scattering,[3,4] µSR,[5] specific heat,[6] and other conventional methods; the ordered moment increases to a maximum of 0.6 $\mu_B$ for x = 0.05. In contrast, SMAF is convincingly observed only via neutron and magnetic x-ray scattering.[4,7,8] It has been speculated that the SMAF state is not static, but fluctuating in time,[4,9,10] while the LMAF is a conventional statically-ordered AFM state.[4] This view is consistent with recent work[11] which indicates that the Pd-concentration at which $T_c$ is suppressed to 0 K is also the concentration at which the LMAF state is first observed, implying that the presence of magnetic fluctuations plays a crucial role in forming the superconducting state. From these considerations it is clear that an understanding of the effects of Pd-substitution on UPt$_3$ can provide substantial insight into the microscopic origins of superconductivity and magnetism, and their interplay, in this system.

In this work we will present data on the suppression of superconductivity in high quality single crystals of U(Pt$_{1-x}$Pd$_x$)$_3$ by varying the Pd content. This work parallels a recent study of the suppression of superconductivity by defects in single crystals of pure UPt$_3$,[12] and a comparison with that work will show that Pd-substitution causes changes beyond those expected for simple impurity potential scattering.

Samples were grown from starting materials of natural U with a purity of 99.98%, and Pt and Pd with 99.999% purity. Single crystal samples of $U(Pt_{1-x} Pd_x)_3$ with x = 0 and 0.001 were prepared in a mirror furnace via the horizontal floating zone technique. The x = 0.002 crystal was prepared in a tri-arc furnace using the Czochralski technique. $U(Pt_{1-x} Pd_x)_3$ forms in a HCP-like lattice for $x \leq 0.15$.[13] Samples were subsequently annealed at 950 °C for four days in high vacuum with a uranium getter, followed by a two-day cooldown.[4] Pd is isoelectronic with Pt, and because the ionic radii of the two are nearly equal, the fractional unit cell volume change associated with substituting 1% of Pt with Pd is less than $10^{-4}$, and is caused primarily by a reduction of the c/a ratio. A similar reduction in volume would result from the application of a hydrostatic pressure of about 0.2 kbar, although in that case the c/a ratio increases due to the anisotropic compressibility.[13]

The pure $UPt_3$ single crystal has a room-temperature to residual resistivity ratio of 780 for current flow along the c-axis, in good agreement with data presented in Ref. [12] for samples annealed under similar conditions. Based on the impurity levels of our starting materials, we estimate our nominally pure $UPt_3$ crystal to have a residual resistivity of about 0.2 $\mu\Omega$ cm, and this is consistent with the results to be presented. In addition to the resistivity data presented in this work, all the crystals have been characterized through x-ray Laue patterns, neutron-scattering[3,4] and specific heat measurements.[2] The high quality of the crystals is evidenced by the well-defined double-superconducting transitions observed for all three annealed crystals.

We have cut several samples from the three single crystals described above. For a given concentration and orientation, there are small deviations in the resistive parameters of the samples. Studies on polycrystals[1,11] show that the residual resistivity depends linearly on Pd-concentration for x < 0.01, and we will assume here that the sample-to-sample variations for a given nominal Pd-concentration are primarily caused by slight variations in the actual Pd-concentrations.

Electrical leads were soldered to the samples. Resistance was measured by a four-terminal AC bridge method, with excitation currents of 100 µA at a frequency of 15 Hz. The current dependence of the resistive parameters was measured to ensure that no self-heating of the samples occurred. Measurements were carried out in a $^3$He refrigerator with a calibrated $RuO_2$ thermometer in Amsterdam; additional measurements were then carried out on some of the same samples in a $^3$He refrigerator with a commercially calibrated Cernox[14] thermometer at Boston College, and no change quantitative in the results was observed. The resistivity is calculated from the resistance by normalizing our room temperature values to 238 µΩ cm (current along the a-axis) and 132 µΩ cm (current along the c-axis), as measured for single crystal $UPt_3$,[15] and we assume that the room temperature values are unaffected by the addition of small amounts of Pd.

In Figure 1 we show typical resistivity versus temperature curves for three samples with nominal Pd concentrations of x = 0, 0.001, and 0.002 and for current directed along the a and c axis. The data above $T_c$ are can be described by the Fermi liquid theory expression

$$\rho_i = \rho_{0,i} + A_i T^2 \quad (1)$$

for temperatures below about 1 K; the subscript i (=a,c) denotes the current flow axis. The constant and quadratic terms represent the elastic and inelastic scattering contributions, respectively. In Figure 2 we plot the results for the inelastic scattering coefficient $A_i$ vs. $\rho_{0,i}$, extracted from data in Figure 1 as well as from data for different samples cut from the same crystals. The $A_i$ increase linearly with $\rho_0$, and the magnitude of the increase is significantly larger than that observed for substituting Y or Th for U.[1] In Ref. [12], the $A_i$ values were approximately constant, although studied over a small range of $\rho_0$ values. As $\rho_0 \to 0$, we find a ratio of $A_a/A_c = 2.70 \pm 0.13$, in good agreement with Ref [12]. Nonetheless, the strong variation of the $A_i$ with increasing $\rho_0$ clearly indicates that *Pd-substitution produces effects beyond simple impurity or defect scattering*, which would leave the $A_i$ unchanged.

The increase of the $A_i$ values with Pd-substitution for these samples is accompanied by an increase[2] in the electronic coefficient of specific heat $\gamma$ such that the Kadowaki-Woods ratio[16] $A_i/\gamma^2$ remains constant within the experimental error of roughly 10%. This is strong evidence that the increase in the $A_i$ is related to bulk properties which dominate the thermodynamics of the system.

In Figure 3 we show the results for $T_c$, determined by a 50% drop in the resistance relative to the extrapolated normal-state resistance, as a function of $\rho_0$. The data are well-described by a linear suppression. A similar linear relationship was obtained in Ref [12] (also shown in Figure 3), as well as for polycrystals of the type $U_{1-x}M_xPt_3$ with M representing a variety of rare earth and other ions.[17] Our rate of suppression with residual resistivity with current along the c-axis, $-dT_c/d\rho_{0,c}$, is $166 \pm 10$ mK $\mu\Omega^{-1}$ cm$^{-1}$, while the corresponding rate for the a-axis data is $65 \pm 5$ mK $\mu\Omega^{-1}$ cm$^{-1}$; both are much larger than the corresponding rates reported in Ref. [12], 96 mK $\mu\Omega^{-1}$ cm$^{-1}$ and 27 mK $\mu\Omega^{-1}$ cm$^{-1}$, respectively.

Extrapolation of our data to $\rho_0 = 0$ yields an "intrinsic" $T_c (\rho_0 = 0) = 578 \pm 6$ mK, significantly higher than the value $563 \pm 5$ mK obtained in Ref [12]. It is possible that this discrepancy arises from the fact that $T_c$ may vary non-linearly with $\rho_0$, and we must extrapolate from higher values of $\rho_0$ compared to those of Ref [12]. However, we have measured a separate series of polycrystalline samples of pure UPt$_3$, with residual resistivities in the range 0.17 – 0.30 $\mu\Omega$ cm, and we have found values of $T_c$ in excess of 570 mK,[11,18] indicating that the higher value may indeed be correct.

As a check on our data, which are taken over a limited range of $\rho_0$ values, we again refer to data taken by us[11] on *polycrystalline* U(Pt$_{1-x}$ Pd$_x$)$_3$ with $0 \leq x \leq 0.006$. In that work, we also found a linear suppression of $T_c$ with $\rho_0$ for values up to 5 $\mu\Omega$ cm. The intrinsic $T_c(\rho_0=0) = 580 \pm 8$ mK, and the rate of suppression is $- dT_c /d\rho_0 = 77 \pm 4$ mK $\mu\Omega^{-1}$ cm$^{-1}$. Clearly the intrinsic $T_c$ is equivalent to our single crystal result. The rate of suppression should be compared to the a-axis value for the single crystals due to preferential alignment during the arc-melting process.[1] The result is very close to the single crystal value of 65 mK $\mu\Omega^{-1}$ cm$^{-1}$, the slightly higher value presumably due to the imperfect alignment in the a-b plane.

Finally, we show in Figure 4 the details of the resistive superconducting transitions for different Pd concentrations. Interestingly, while the width of the resistive transition (as determined by a 10% -90% criterion) is unaffected by Pd concentrations up to 0.002 for current along the c-axis, for current along the a-axis the width increases from 7 mK to 19 mK.

As discussed in Ref. [12], if the inelastic scattering probabilities are isotropic, the anisotropies in the values of $\rho_{0,i}$ and $A_i$ can be used to extract the anisotropy of the elastic scattering times $\tau_i$

$$\tau_c/\tau_a = \frac{\rho_{0,a} A_c}{\rho_{0,c} A_a} \quad . \quad (2)$$

The ratio $\rho_{0,c}/\rho_{0,a} = 0.39$, as found from the slopes of Fig. 3. Because the $A_i$ vary with Pd-content we use the ratio extrapolated to $\rho_0 = 0$. Thus we find

$$\tau_c/\tau_a = 1.0 \pm 0.2 ,$$

to be compared to the results of Ref. [12], $\tau_c/\tau_a = 1.3 \pm 0.1$. In terms of the calculations presented in Ref. [12] for an anisotropic unconventional superconductor with $E_{1g}$ or $E_{2u}$ pairing symmetry, our results imply that only s-wave impurity scattering is significant in this system. However, the relatively large error in our result makes an interpretation difficult.

The $T_c$ suppression rates with residual resistivity are significantly higher for Pd-substitution than for suppression by defects. It is also higher than for substitutions on the U-site,[17,19] where the rate for various substitutions $U_{0.997}M_{0.003}Pt_3$ is in the range 40-50 mK $\mu\Omega^{-1}$ cm$^{-1}$, a value consistent with the defect suppression rate. Once again it is found that Pd-substitution is anomalous. There is relatively little data for substitutions other than Pd on the Pt-site. Ni-substitution produced multiple phases.[19] Limited studies on polycrystalline samples show that Au-subsitution suppresses $T_c$ at a rate similar to the U-site substitutions, while Ir-substitution has a value of $-dT_c/d\rho_0$ that may be comparable to that for Pd-substitution.[20] No data exists regarding its effect on the $A_i$.

As given in Ref. [17], the shift of $T_c$ from its zero-impurity value, $\Delta T_c$, can be written as the sum of impurity potential and spin-flip scattering terms. It has been clearly demonstrated that both magnetic and non-magnetic impurities suppress superconductivity at the same rate with respect to $\rho_0$ in UPt$_3$[17,19]. The lack of a spin-flip scattering contribution indicates parallel-spin electron pairing, a conclusion also drawn from recent NMR measurements.[10] Thus we believe that magnetic pair-breaking is not playing any significant role in the suppression of $T_c$.

Our results suggest that the enhanced value of $-dT_c/d\rho_{0,i}$ observed for Pd-substitution is directly related to the increase in the $A_i$. This claim is supported by specific heat measurements on UPt$_3$[21] which show that applying stress along the c-axis reduces $T_c$. While the $A_i$ coefficients were not measured in that work, it was found that the normal-state electronic coefficient of specific heat increased. Assuming a constant Kadowaki-Woods ratio, this implies that the $A_i$ increase as $T_c$ decreases, as for our results. It has also been reported that pulling whiskers of UPt$_3$ along the c-axis causes $T_c$ to increase, and the A-coefficient to decrease.[22] Because Pd has the effect of reducing the c/a ratio, these results indicate that c/a may play an important role in forming the superconducting state, as it does for magnetism[4] in UPt$_3$.

Our correlation between $-dT_c/d\rho_{0,i}$ and $A_i$ can be understood qualitatively if we assume that in addition to impurity potential scattering, there is a contribution to pair-breaking from inelastic scattering processes. For example, Millis, Sachdev, and Varma[23] described how electron scattering from low-frequency antiferromagnetic fluctuations could lead to pair-breaking in anisotropic, spin-fluctuation-mediated superconductors (calculated for s- and d-wave pairing only). Assuming that the low-temperature resistivity of UPt$_3$ results from spin-fluctuating Fermi liquid behavior,[24] then the $A_i \propto T_{sf}^{-2}$, where $T_{sf}$ is the characteristic spin-fluctuation temperature ($\approx$ 20 K in UPt$_3$[25]). Pd-substitution increases $A_i$, therefore decreasing $T_{sf}$. This results in an increase in the low-temperature spin-fluctuations and enhanced inelastic scattering. We note that the increase in spin-fluctuations with increasing Pd-content is also consistent with the approach to static magnetic ordering (the LMAF phase) when x >0.007.

This qualitative explanation has some difficulties. For example, increasing hydrostatic pressure reduces both $T_c$ and the $A_i$.[25] However, it must be remembered that the magnetic fluctuation spectrum of UPt$_3$ is complex,[7, 26-28] and so the value of the inelastic scattering coefficient may contain many contributions in addition to fluctuations resulting in the pair-breaking inelastic scattering processes. To fully characterize these relationships would require a complete comparison of the magnetic fluctuation spectrum for U(Pt$_{1-x}$ Pd$_x$)$_3$ with moderate amounts of Pd substituted (e.g., x = 0.005) with pure UPt$_3$ at ambient and high-pressure. Such a study is beyond the scope of this Brief Report.

Finally, we comment on the anisotropic broadening of the superconducting transition shown in Fig. 4. The simplest magnetic structure in agreement with neutron scattering results for the SMAF phase is single-**q** with the local magnetic moments aligned parallel or anti-parallel to the a-axis.[29] Thus it is natural to associate the broadening of $T_c$ with the measured increase in the SMAF ordered moment which occurs for Pd-substitution. There are three equivalent magnetic domains in the a-b plane for this structure and so it is possible that the broadening for I//a-axis results from scattering at domain walls. This does not occur for current directed along the c-axis as all three magnetic domains are equivalent along the c-axis.

In conclusion, we have measured the suppression of superconductivity in high quality single crystals of UPt$_3$ by substitution of Pd for Pt. The rate of suppression with residual resistivity $-dT_c/d\rho_0$ is larger than similar rates caused by defects or other impurity substitutions, and may be associated pair-breaking by inelastic scattering processes resulting from enhanced spin fluctuations.

The work in Amsterdam was supported through Dutch Foundation for Fundamental Research on Materials ("Stichting" FOM). Work at Boston College was supported through Research Corporation Grant RA0246. MJG and AdV acknowledge support through NATO Collaborative Research Grant CGR11096. MJG would like to thank colleagues at the Van der Waals – Zeeman Institute for their generous hospitality during his recent sabbatical leave.

**Figure Captions**

Fig. 1  Resistivity plotted versus $T^2$ for $U(Pt_{1-x} Pd_x)_3$ with nominal concentrations of x = 0, 0.001, and 0.002 for current (a) along the a-axis, and (b) along the c-axis.  Lines are linear fits.

Fig. 2  Inelastic scattering coefficient plotted versus residual resistivity for current along a-axis and c-axis.  Solid circles are the data, and solid lines are linear fits.  The open circles are data taken from Ref. [11] .

Fig. 3  Superconducting transition temperature plotted versus residual resistivity for current along a-axis and c-axis.  Solid circles are the data, and solid lines are linear fits.  The open circles and dashed lines are data and linear fits, respectively, taken from Ref. [11] .

Fig. 4  Resistivity divided by the (extrapolated) normal state resistivity plotted versus $T-T_c$ in the vicinity of $T_c$ for $U(Pt_{1-x} Pd_x)_3$ with nominal concentrations of x = 0, 0.001, and 0.002 for current (a) along the a-axis, and (b) along the c-axis.